\documentclass[aps, prd, amsmath, floats, floatfix, twocolumn, nofootinbib]{revtex4-1}
\usepackage{graphicx}
\usepackage{comment}
\usepackage{latexsym}
\usepackage{color}
\usepackage{ dsfont }
\usepackage{amsfonts,amsbsy}
\usepackage[utf8]{inputenc}
\usepackage[mathscr]{euscript}
\usepackage{ amssymb }
\usepackage{amsthm}
\usepackage{amsmath}
\usepackage[T1]{fontenc}
\usepackage[utf8]{inputenc}
\usepackage{mathtools}
\usepackage{physics}
\usepackage[utf8]{inputenc}
\usepackage[english]{babel}

\newcommand{\be}{\begin{eqnarray}}
\newcommand{\ee}{\end{eqnarray}}

\theoremstyle{definition}

\begin{document}

\title{Entanglement swapping via black holes: restoring predictability}

\author{Ali Akil}
\email{aakil@connect.ust.hk}

\author{Oscar  Dahlsten}
\email{dahlsten@sustech.edu.cn}

\author{Leonardo Modesto}
\email{lmodesto@sustech.edu.cn}

\affiliation{$^{*\, \dagger \, \ddagger}$Department of Physics, Southern University of Science and Technology (SUSTech), Shenzhen 518055, China}

\affiliation{$^{*}$ Department of Physics, The Hong Kong University of Science and Technology, Clear Water Bay, Kowloon, Hong Kong, China.\\
Jockey Club Institute for Advanced Study, The Hong Kong University of Science and Technology, Clear Water Bay, Kowloon, Hong Kong, China}

\affiliation{$^\dagger$Institute for Quantum Science and Engineering (SUSTech), Shenzhen 518055, China}

\affiliation{$^\dagger$Wolfson College, University of Oxford, Linton Rd, Oxford OX2 6UD, United Kingdom}

\affiliation{$^\dagger$London Institute for Mathematical Sciences,
35a South Street, Mayfair, London, W1K 2XF, United Kingdom}

\begin{abstract}
Black holes evaporate into radiation which may hold little information about the black hole matter, according to Hawking’s model. Pure quantum states forming the black hole become mixed, violating the quantum theory postulate that isolated systems evolve unitarily. We track the information encoded in the radiation particles throughout the evaporation process. Conditional upon full evaporation, entanglement between inside and outside modes is swapped to the outside. The  outside radiation is then pure, as well as one-to-one with the initial state forming the black hole, preserving information.
\end{abstract}

\maketitle

{\em Introduction --}Hawking's model for black hole evaporation implies a many-to-one mapping of initial pure states to a more random mixed state \cite{Hawking74, Hawking76}. This has been dubbed a `loss of predictability'. 
In Hawking's model, pairs of particles are created from the vacuum near the event horizon: one of these (having negative energy\footnote{Inside a Schwarzschild black hole, the time coordinate and the spatial radial coordinate interchange their roles. Therefore, the energy and momentum change their roles, too, allowing for a well-defined negative energy (actually, spatial momentum) inside the black hole. This fact is at the heart of the Hawking process for black holes' evaporation.}) falls into the black hole and the other flies away to future infinity ($I^+$).
The particle of negative energy falling towards the black hole will eventually meet the black hole's matter and annihilate, causing the black hole mass to decrease \cite{Hawking74, Hawking76,  Susskind:2005js, Fabbri}. As time passes, increasingly more particles are annihilated and the black hole eventually evaporates.  The particle pairs created at the event horizon are, before the annihilation, in the following state \cite{Fabbri}, 
\be 
 \ket{\Psi} = \bigotimes_{\omega > 0} c_\omega \sum_{N_\omega=0} e^{-\frac{ N \pi \omega} { \kappa} } \ket{ N_\omega} ^{\rm out} \otimes \ket{ N_\omega}^{\rm int}, 
\label{Main}
 \ee
where $c_\omega \equiv \sqrt{1- e^{- 2 \pi \omega/\kappa}}$ is a normalization factor, 
$N_\omega$ is the number of particle pairs of energy $\omega$, while ``int'' and ``out'' label the Hilbert spaces for the particles falling inside the black hole and those escaping to the future infinity, respectively \cite{Fabbri}. The state $\ket{\Psi}$ of Eq.(\ref{Main}) is pure with the ``int'' modes inside the black hole being correlated with the ``out'' modes. After the annihilation, the ``int'' modes and the black hole are in a vacuum state, and the ``out'' modes  are the non-trivial remnants. Thus according to this model of evaporation the black hole's initial state is finally mapped to $\mathrm{Tr}_{\mathrm{int}} (\ket{\Psi}\bra{\Psi}$) \cite{HawkingInfo}. As this state is mixed and even independent of the black hole's initial state the evolution is non-unitary. However, closed quantum systems are expected to evolve unitarily \cite{Nielsen}. These two contradicting statements are the so-called black hole information paradox.   

Multiple interesting approaches have been proposed in connection with resolving the paradox. Hawking's semi-classical approximations were questioned \cite{Page0}. Quantum gravitational corrections to General Relativity were proposed that would leave a ``remnant'' upon evaporation \cite{QGbook, Modesto:2010uh, Leonardo23, Rovelli, Perez}. Modifying quantum theory through nonlinear effects, non-violent nonlocal effects, and generalized probabilistic theories was considered \cite{Kapuccino, Maldacena, Oscar, Lloyd, Giddings}. Of particular relevance here is that Page noted the possibility of quantum correlations between the early emitted radiation and the late radiation such that the total radiation state could be pure, but subsystems of the radiation mixed \cite{Page, Page2}. However, that scenario was argued to incompatible with the monogamy of entanglement \cite{Coffman, Mathur}. The ``out'' modes and ``int'' modes would be strongly entangled yet the early radiated particles would also be strongly entangled with the late radiated particles. New physical principles and phenomena like ``Complementarity'' and entanglement-breaking (and possibly equivalence-principle violating) ``Firewalls'' have also been considered \cite{Complementarity, AMPS, Braunstein}. Models with global unitary dynamics disentangling the radiation from the black hole, which may come with the price of non-causal signalling, have also been proposed \cite{Verlinde, Sabine}.

Here we propose an alternative approach. Our main result is a derivation, showing that upon evaporation, the entanglement and the information is transferred to the outside modes. We assume essentially Hawking's model. The annihilation process inside the black hole induces entanglement swapping \cite{Zei1, Zei2,ex1, ex2, ex3, ex4, ex5, ex6, ex7}. Conditional on the (full) annihilation of the black hole, the outside radiation is indeed entangled and in a pure state. There is no information loss in that the final radiation state is one-to-one with the initial black hole state. No basic principles are violated because we use a standard black hole model and standard quantum theory, and in particular there is no signalling for the same reason that entanglement swapping is not signalling.  Another attractive feature of the model is that the entanglement swapping appears naturally when modelling the annihilation explicitly as opposed to being postulated. 

We proceed as follows. We describe the evaporation model. We show firstly how the entanglement between the ``out'' and ``int'' modes gets swapped to entanglement within the ``out'' modes upon annihilation. We then show that there is no information loss, in that the initial state of the black hole matter is in a one-to-one correspondence with the state of the radiation after full evaporation. In the supplementary material, we briefly review entanglement swapping and why it does not allow signalling, then we present generalizations of the toy model used in the main body, and finally, we discuss the role of the singularity in this type of evaporation model and how different proposed modifications impact our conclusions.

{\em Entanglement swapping in Black Holes --} \label{SwappBH}
The Hawking radiation state in (\ref{Main}) describes all the radiated particles, 
 but for a better exposure and analysis of the problem we can focus on one pair being created near the event horizon. Therefore, the state (\ref{Main}) simplifies to\footnote{Considering one pair means $\ket{\Psi}$ is restrited to the subspace where $\sum_{\omega }N_{\omega}=1$. Tensor products (in the first quantisation frequency picture) of such pairs form the more general state $\ket{\Psi}$ (in the second quantisation picture) so this is a natural elementary unit and calculations for this case readily generalise, as we show in the supplementary material. }: 
\be 
\label{Main2}
\hspace{-0.4cm}
 \ket{\psi} = \mathcal N \sum_\omega  e^{-\frac{  \pi \omega} { \kappa} } 
 \ket{ \omega} ^{\rm out} \otimes \ket{-\omega}^{\rm int},
 \ee
where $\mathcal N$ is a normalization factor, which, being a function of products of all possible $\omega$'s, does not depend on $\omega$. 
Now instead of tracing out the ``int'' modes, we carefully track the entanglement as the infalling particles annihilate with the black hole matter. In general, the black hole has some mass $M$ as a result of the gravitational collapse of a large number of entangled particles in a pure state (we will also consider the case of particles that are entangled with nothing else). However, in this paragraph, for the sake of presentational clarity, we consider only one entangled pair inside the BH described at a time, and give the general case in the supplementary material. Thus we now consider the following state inside the black hole:
\be 
\ket{\phi} =\sum_ {\omega'} f(\omega') \ket{\omega'}_{\rm A} \otimes \ket{\omega'}_{\rm B } \, .
\label{BHp}
\ee
Note that our argument does not rely on the specific form of $f(\omega)$. The initial state is given by the tensor product of (\ref{Main2}) and (\ref{BHp}),  
namely%
\be
\!\!\! | i \rangle  && = | \psi \rangle \otimes | \phi \rangle  \nonumber \\
&& =\mathcal N \sum_{\omega, \omega'}  f( \omega' ) {\rm e}^{-\frac{  \pi \omega} { \kappa}} \ket{ \omega} ^{\rm out}  \ket{-\omega}^{\rm int}  \ket{\omega'}_{\rm A}  \ket{\omega'}_{\rm B } \, . 
\label{IN}
\ee
If the incident negative energy particle $| - \omega \rangle^{\rm int}$ interacts with the particle A of energy $\omega'$, the interaction outcome is labelled by the state $\ket{\omega'  -\omega}^{\rm int}_A$ . Therefore, after the interaction has occurred, the state is:
\be
&& \hspace{-0.95cm}  | f_1 \rangle = \mathcal N \sum_{\omega', \omega}  f( \omega' ) e^{-\frac{  \pi \omega} { \kappa}} \ket{ \omega} ^{\rm out}  \ket{\omega'  -\omega}^{\rm int}_{\rm A}  \ket{\omega'}_{\rm B } \label{fg} 
\label{FIN}
\ee
Now $| f_1 \rangle $ is our new state and the mass of the black hole is reduced to $M - \omega$. Note that there is a sum over the $\omega$'s, so the black hole is in a superposition of energy eigenstates. Moverover, we ended up with three particles (or subsystems) entangled\footnote{In our treatment of the annihilation process we have labeled the states with their energies $\ket{\omega}$, although $\omega$ does not fully specify the state. We omitted the momentum label $p$ because it is not a conserved quantity and for notational simplicity.}.
\begin{figure}
\begin{center}
\hspace{0cm}
\includegraphics[width=\linewidth]{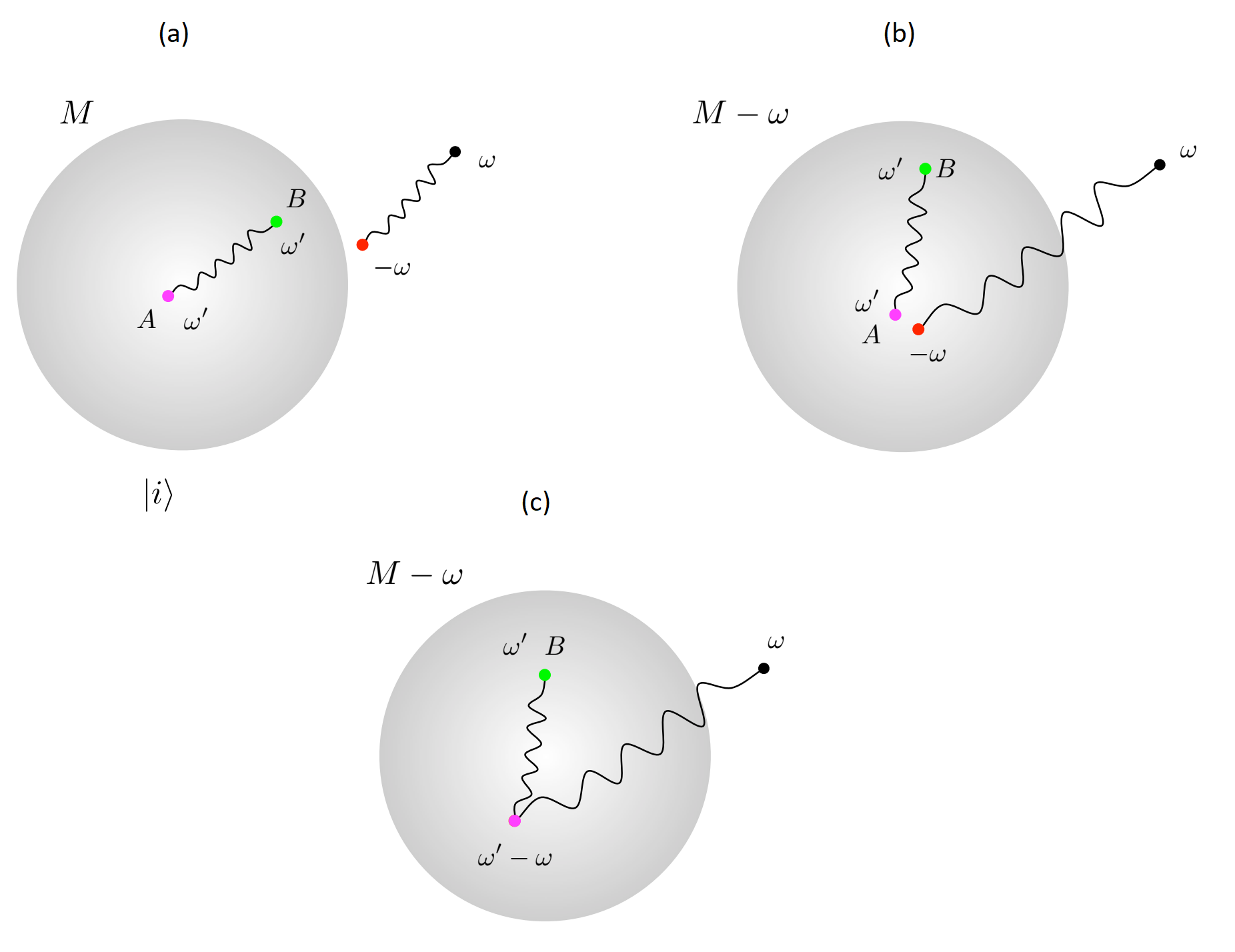}
 \label{f1b}
 \end{center}
\caption{{\em Annihilation ---} There are two entangled pairs, one internal to the black hole and another created at the horizon, a Hawking pair ( Fig. 1.a). the negative energy particle falls into the black hole (Fig. 1.b). It interacts with the particle A (Fig. 1.c).}
\end{figure}
\begin{figure}
\begin{center}
\includegraphics[width=\linewidth]{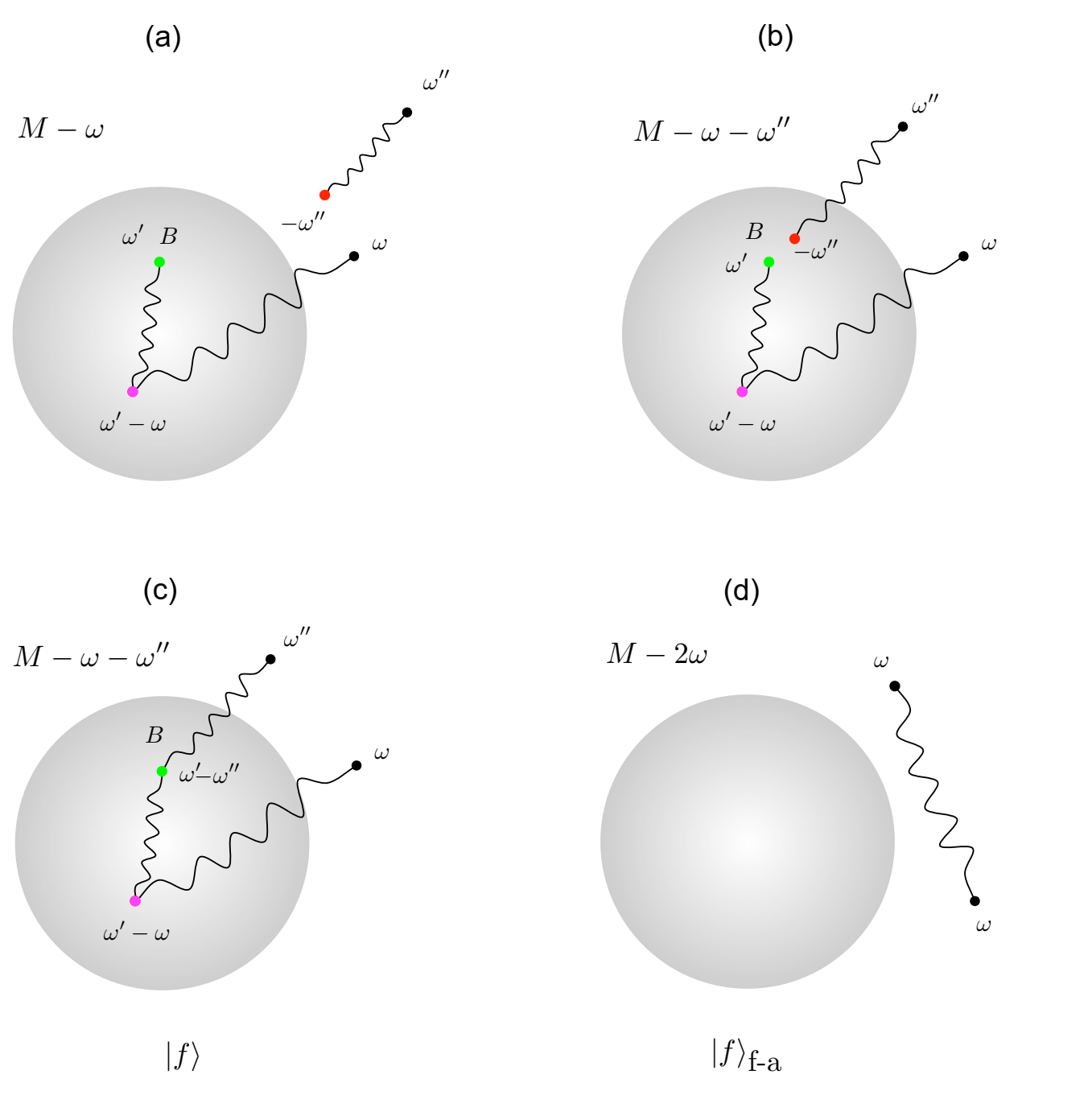}
 \label{fig2finA}
 \end{center}
\caption{{\em Entanglement transferred outside ---} Following from Fig. 1.c, a second pair is created near the event horizon (Fig. 2.a). In Fig. 2.b, the particle with negative energy $-\omega''$ crosses the horizon and scatters with particle 
B with energy $\omega'$ in Fig. 2.c. If we have full annihilation inside the black hole, namely $\omega'' = \omega' = \omega$, then we end up with the situation shown in Fig. 2.d when the ``out'' particles are entangled and the black hole mass is $M-2 \omega$.}
\end{figure} 
Let us now consider a second Hawking pair created near the Event Horizon, namely $\ket{\psi_2}$. Using again (\ref{Main2}) for $\ket{\psi_2}$ and assuming $| f \rangle$ as the initial state, 
the whole system is described by the tensor product $| i' \rangle = | f \rangle \otimes  \ket{\psi_2}$ (see Fig.2.a-b), namely 
\be
\hspace{-0.5cm} 
| i' \rangle =\mathcal N~^2 \!\!\! && \sum_{\omega'', \omega', \omega} f( \omega' )  e^{-  \pi\left(\frac{ \omega } { \kappa}+ \frac{\omega''}{\kappa''}\right)}  \ket{ \omega} ^{\rm out} \nonumber  \\
&&  \otimes \ket{\omega'  -\omega}^{\rm int}_{\rm A}  \ket{\omega'}_{\rm B }  
 \ket{ \omega^{\prime \prime}} ^{\rm out}  \ket{-\omega^{\prime \prime}}^{\rm int}  \, . 
\label{casei-1}
\ee
When the newly created Hawking particle interacts with the particle B, the resulting state is 
\be
\hspace{-0.5cm} 
| f' \rangle =\mathcal N~^2 \! \! \! \sum_{\omega'', \omega', \omega}\!\!\!\! &&  f( \omega' ) e^{-  \pi\left(\frac{ \omega } { \kappa}+ \frac{\omega''}{\kappa''}\right)}   \ket{ \omega} ^{\rm out}  \ket{ \omega^{\prime \prime}} ^{\rm out} \nonumber \\
&&   \otimes \ket{\omega'  -\omega}^{\rm int}_{\rm A}  \ket{\omega'- \omega''}^{\rm int}_{\rm B }  \, . 
\label{casei0}
\ee
The resulting state $| f' \rangle $ consists of two particles inside the black hole partially entangled between each other and with  the two Hawking particles outside (as illustrated in Fig. 2.c). 
Finally, assuming full annihilation of the two particles inside (or in this toy model: full evaporation of the black hole), it is easy to see that one gets 
\be
\hspace{-0.5cm} 
| f_{\rm Evap} \rangle = \mathcal N~^2 \sum_{\omega}  f( \omega )   && e^{-  \pi  \omega  \left(\frac{1} { \kappa}+ \frac{1}{\kappa''(\omega)}\right)} \ket{ \omega} ^{\rm out}  \ket{ \omega} ^{\rm out} \nonumber \\
&&   \otimes \ket{0}^{\rm int}_{\rm A}  \ket{0}^{\rm int}_{\rm B }  \, , 
\label{casei}
\ee
which is clearly an entangled pair outside the black hole (see Fig.2.d). Thus the entanglement of the black hole matter has been swapped to the outside. The pure state (\ref{BHp}) has evolved to a similar pure state (\ref{casei}), regardless of whether you trace out the ``int'' modes.

This model can be readily generalised. As detailed in the supplementary material, the same conclusions follow if the incident Hawking particle scatters to produce more than one particle inside the black hole or interacts with another particle. In addition, more general black hole states, for example, when the inside annihilated particle is not entangled with anything, lead to the same conclusion concerning entanglement and purity.

{\em No Loss Of Information --}
We now focus on the question of information loss in the black hole evaporation process. In particular, we are going to show that information is not lost, in the sense that one can recover the initial state of the black hole given the final state of the radiation upon annihilation and the evolution operator.  
We consider again the black hole and radiation states in the same settings.
We recall the black hole's initial state
\be
\ket{\phi} =\sum_ {\omega'} f(\omega') \ket{\omega'}_{\rm A} \otimes \ket{\omega'}_{\rm B } \, ,
\ee
then, after emission of Hawking particles, we get
\be
\hspace{-0.5cm} 
| f' \rangle =\mathcal N~^2  \!\!\! \sum_{\omega'', \omega', \omega} \!\!\!\!&& f( \omega' ) e^{-  \pi\left(\frac{ \omega } { \kappa}+ \frac{\omega''}{\kappa''}\right)} \ket{ \omega} ^{\rm out}  \ket{ \omega^{\prime \prime}} ^{\rm out} \nonumber \\
&&  \otimes  \ket{\omega'  -\omega}^{\rm int}_{\rm A}  \ket{\omega'- \omega''}^{\rm int}_{\rm B }  \, . 
\ee
Given the final state:
\be
\hspace{-0.5cm} 
| f_{\rm Evap} \rangle =  \mathcal N~^2 \sum_{\omega}  && f( \omega ) e^{-  \pi  \omega  \left(\frac{1} { \kappa}+ \frac{1}{\kappa''(\omega)}\right)} \ket{ \omega} ^{\rm out}  \ket{ \omega} ^{\rm out} \nonumber \\
&&  \otimes  \ket{0}^{\rm int}_{\rm A}  \ket{0}^{\rm int}_{\rm B }  \, ,
\label{evaporated}
\ee
and the evolution operator that takes the initial black hole matter state to the final fully evaporated state
\be \label{evolution}
 O=\mathcal N~^2 && \sum_\omega {\rm e}^{- \pi \omega \left(\frac{1}{\kappa (\omega)}+\frac{1}{\kappa'' (\omega) }\right)} \nonumber \\
&& \!\!\!\! \!\!\! \ket{\omega_{\rm out},\omega_{\rm out},0_{\rm int},0_{\rm int}}\bra{0_{\rm out},0_{\rm out},\omega_{\rm int},\omega_{\rm int}}, 
 \ee
we would like to see whether we can reconstruct the initial state.
Indeed it is quite simple to show that we can invert the evolution operator $O$ and thus reconstruct the initial state. 
First of all, one can extract the relative factors in $O$ from the measurement of $\omega$, namely 
\be
\mathcal N~^2 {\rm e}^{- \pi \omega \left(\frac{1}{\kappa (\omega)}+\frac{1}{\kappa'' (\omega) }\right)}= \bra{0,0,\omega,\omega} O \ket{0,0,\omega,\omega}.
\ee
Then one can use the final state (\ref{evaporated}) to construct the initial one,
\begin{eqnarray*}
\hspace{-0.3cm} 
|{\rm initial} \rangle \!\!&=& \!\!   \sum_{\omega} \left( \bra{0,0,\omega,\omega} O \ket{0,0,\omega,\omega} \right)^{-1}  \mathcal N~^2 f( \omega ) \nonumber \\
&& e^{-  \pi  \omega  \left(\frac{1} { \kappa}+ \frac{1}{\kappa''(\omega)}\right)} \ket{ \omega} ^{\rm out}  \ket{ \omega} ^{\rm out}  \ket{0}^{\rm int}_{\rm A}  \ket{0}^{\rm int}_{\rm B }\\  \\
 \!\!\! &=&  \!\! \sum_ {\omega'} f(\omega') \ket{\omega'}_{\rm A} \otimes \ket{\omega'}_{\rm B }
 \, . 
\end{eqnarray*}
Thus, knowing the final state for radiation, and knowing the evolution operator eq. (\ref{evolution}), one can easily reconstruct the initial black hole matter state. Therefore, no information is lost after full evaporation.

{\em The singularity issue --}
One may question whether the existence of a singularity affects the outcome of our analysis. There are reasons to believe that our solution of the information loss problem seems to work regardless of whether the spacetime is singular or singularity-free \cite{Modesto:2010uh,Leonardo23, Nicolini, Leonardo, Leonardo2, tHooft:2016uxd, Bambi:2016yne,Bambi:2016wdn,Brustein}.  Therefore, 
in this work we do not intend to address and/or solve the singularity issue, but only comment on it. Indeed, our result seems to be valid for any black hole whose geometry allows interactions between the black hole's matter and the infalling Hawking particles. 
On the other hand, for black holes where these particles do not interact, there is no reason for the evaporation to happen, as we are going to explain. In the previous sections we never mention the spacetime singularity issue at $r=0$. Our analysis is based on the natural assumption that particles inside the black hole are annihilated by the Hawking negative energy particles. 

Now let us make some comments on the singular black hole case in particular. As long as the ``int'' particles interact with the BH matter that has not reached $r=0$ yet, the dynamics (the S-matrix) is well defined and the scattering takes place without violating unitarity. On the other hand, for $v > v_s$ the ``int'' particles probably annihilate with matter particles that have already reached the singularity (see Fig.5). In this paper as well as most others in the literature, it is assumed that the annihilation takes place regardless of the singularity\footnote{As proved in the paper \cite{Balasin}, 
 titled ``The energy-momentum tensor of a black hole, or what curves the Schwarzschild geometry?'', 
the source of the Ricci flat solutions (in vacuum) has a well defined meaning in the space of distributions and the energy-momentum tensor is proportional to the Dirac's delta, namely ${\bf T} \propto M \delta({\bf r})$ (this is also proved in many other textbooks like Landau-Lifshitz, etc). After the black hole formation, the matter is localized at $r=0$ and can be reached in finite time (or finite value of the affine parameter in the massless case) by the Hawking ``int'' particles. Therefore, all the ``int'' particles annihilate for $r>0$ in the first stage of the evaporation process or in $r=0$ afterwards to finally end up with zero Bondi-Sachs mass. Notice, that if there was no source at $r=0$ then the spacetime would be Minkowski and not Schwarzschild.}. 
Therefore, we are entitled to believe that if a singular black hole ever evaporates, then entanglement is transferred to (and/or from) the matter at the singularity. 
On the other hand, if there is no annihilation at the singularity there seems to be no reason 
do not have evaporation and thus any information loss problem because there are correlations between the matter inside and the particles outside the black hole, that keep the state of the whole system pure. This eventuality will be surely studied in the future but it does not affect the universality of the content and claims in this paper.
It is worth being stressed that the absence of local (or non-local) interactions between Hawking ``int'' particles and the matter at the singularity implies that there is no black hole evaporation at all, contrary to what is commonly stated
Furthermore, we do not have any information loss problem because the matter would still be there, and the Hawking particles would still be there too. Indeed, the whole information loss business is based on the assumption that the black hole completely (or nearly) evaporates and most of the mass evaporates after the creation of the singularity. 
If we question the interaction of the ``int'' particles with the singularity then we cannot trust the black hole evaporation after the instant $v_s$. However, at this stage of the evaporation process the black hole retains most of its mass, which is enormously bigger than the Planck mass. Why in such semiclassical regime should we not believe in the black hole evaporation? We do not want to address this question in this paper, but we only want to point out that our resolution of the information loss problem is based on very reasonable and common assumptions.

Finally, in any singularity-free black hole our proof is {\em a priori} expected to apply and there is no information loss problem because in this case the spacetime is geodesically complete and the needed interactions for $v > v_s$ can happen smoothly. 
In a future project, we will carefully work out which black hole geometries allow our process of entanglement transfer and which ones (if any) do not. This analysis could eventually 
 support some classical or quantum gravitational theories over some others. 

{\em Comments and Conclusions --}
Let us here summarize our result and make some comments on the distinction relative to the standard Hawking model analysis of the information loss problem. In our approach, when there is annihilation the state is updated to reflect that. This concerns a common effect in quantum teleportation, entanglement swapping and remote state transfer. A state can be highly mixed if one does not update it for a given outcome, but conditional on the given outcome it can be pure, with associated entanglement transfer. According to our analysis, conditional upon annihilation, entanglement is in fact transferred outside the black hole, so that the outside state is pure and one-to-one with the initial state of the black hole.  The standard analysis of the outside state in the Hawking model, in contrast, does not update the state according to the assumption that the annihilation has taken place and does not track the entanglement transfer upon annihilation. The outside state then remains invariant throughout the annihilation, in particular being mixed.  

We emphasize that we did not postulate entanglement swapping, it is actually the outcome of our computation, only assuming full evaporation,  energy conservation, and interaction between Hawking infalling particles and the black hole matter. Moreover we have used the standard formalism of QFT in curved spacetime without postulating new principles/phenomena or violations of fundamental principles like non-signalling, nor is there a violation of the monogamy principle. 

\section*{Acknowledgments}
The authors want to thank Malcolm Perry, Juan Maldacena, Alejandro Perez, and Simone Speziale for very valuable discussions. Ali especially thanks Tong Xi for the endless valuable conversations. Leonardo and Oscar owe special thanks to Daniel Terno, a visiting professor at SUSTech, for sharing his thoughts and suggestions about our work and the paradox in general.


\section*{Supplementary Materials}
\label{sec:supp}

\setcounter{section}{0}

\pagenumbering{arabic}\renewcommand{\thepage}{S.\arabic{page}}

\subsection*{Preliminaries: Entanglement swapping}

Here, we briefly introduce the entanglement swapping phenomenon between two EPR pairs.
We consider two entangled pairs 
$(A,V_1)$ and $(B, V_2)$, each in an antisymmetric polarization-entangled Bell singlet state. Therefore, the state of the entire system is
\be \label{Swapping}
\hspace{-0.5cm}
\ket{\Psi} = \Big(\ket{0_{\rm A} 1_{{\rm V}_1}} - \ket{1_{\rm A} 0_{{{\rm V}_1}}}\Big) \otimes \Big(\ket{0_{\rm B} 1_{{{\rm V}_2}}} - \ket{1_{\rm B} 0_{{{\rm V}_2}}}\Big) .
\ee
Let particle A be with Alice and B with Bob, while Victor keeps $V_1$ and $V_2$. Now, if Victor projects his particles onto a Bell state, they become entangled. At the same time, the particles (A,B) become entangled, according to victor, despite there being absolutely no communication. To see that, we write Eq. (\ref{Swapping}) as
\be \hspace{-0.5cm}
\ket{\Psi}&=& \ket{ 1_{\rm V_1} 1_{\rm V_2}} \ket{0_{\rm A} 0_{\rm B}} - \ket{1_{\rm V_1} 0_{\rm V_2}} \ket{ 0_{ \rm A} 1_{\rm B} } \nonumber \\
&-& \ket{0_{\rm V_1} 1_{\rm V_2} } \ket{1_{\rm A} 0_{\rm B} } + \ket{0_{\rm V_1} 0_{\rm V_2} } \ket{1_{\rm A} 1_{\rm B}} . 
\ee
We then let Victor project his particles on (as an example) the state:
\be
\Big(\ket{ 0_{\rm V_1} 1_{\rm V_2}} - \ket{1_{\rm V_1} 0_{\rm V_2}}\Big)\Big(\bra{ 0_{\rm V_1} 1_{\rm V_2}} - \bra{1_{\rm V_1} 0_{\rm V_2}}\Big) \ .
\label{projV}
\ee
Therefore, the final state reads:
\be \hspace{-0.5cm} \label{monogamy}
\ket{\Psi} = \Big(\ket{0_{\rm V_1} 1_{{\rm V}_2}} -\ket{1_{\rm V_1} 0_{{\rm V}_2}}\Big) \otimes \Big (\ket{0_{\rm A} 1_{\rm B}} - \ket{1_{\rm A} 0_{\rm B}}\Big) \,
\ee
exactly as claimed above. Moreover, as a consequence of the monogamy principle mentioned above and as is clear in (\ref{monogamy}), the entanglement of A with ${\rm V}_1$ is broken as well as the entanglement of B with ${\rm V}_2$ \cite{Zei1, Zei2}.

\subsection*{Swapping does not allow signaling}
Entanglement swapping does not allow instantaneous signaling because the observer that makes the measurement at side X (here, we consider a system made of two sides, X and Y) cannot control the outcome of the measurement. More concretely, one can consider a bipartite system XY described by the density matrix $\rho_{\rm XY}$.
An observer can make measurements on X with different possible outcomes described by the following set of projections: 
\be
\Big\{  \ket 1 \bra 1 _{\rm X}, \dots ,  \ket d \bra d _{\rm X} \Big\} .
\ee
If the observer knows the measurement outcome at A, then the sub-normalized post-measurement state is
\be 
{\rho'_{\rm XY}} \!^{(i)} = \Big( \ket i \bra i_{\rm X} \otimes I_{\rm Y} \Big) \rho_{\rm XY} \Big( \ket i \bra i_{\rm X} \otimes I_{\rm Y} \Big). \ee
However, if one does not know the measurement outcome, then they must sum over all possible outcomes, and the post-measurement state will be
\be 
\rho''_{\rm XY} = \sum_{i} \Big( \ket i \bra i_{\rm X} \otimes I_{\rm Y} \Big) \rho_{\rm XY} \Big( \ket i \bra i_{\rm X} \otimes I_{\rm Y} \Big) 
\ee
The density matrix $\rho'_{\rm XY}\!^{(i)}$ is called a ``conditional density matrix,'' and it is used by an observer that knows the outcome of a measurement on subsystem X to describe the entire system XY.
 Note that 
 \be
 \rho'^{\,(i)}_{\rm Y} = {\rm Tr}_{\rm X} \rho'_{\rm XY}\!^{(i)} \neq \rho_{\rm Y} = {\rm Tr}_{\rm X} \rho_{\rm XY},
 \ee 
 which means that a measurement on X seems to change the state of Y. 
Therefore, one might think that we could send information to Y by making a measurement on X.
However, for an observer that does not know the measurement outcome, the reduced density matrix describing the system Y reads 
 \be \rho''_{\rm Y} && = \Tr_X   \sum_{i} \left( \ket i \bra i_{\rm X} \otimes I_{\rm Y} \right) \rho_{\rm XY} \left( \ket i \bra i_{\rm X} \otimes I_{\rm Y} \right) \nonumber \\ 
&& = \sum_{i,k} \bra{k}_{\rm X} \Big( \ket i \bra i_{\rm X} \otimes I_{\rm Y} \Big) \rho_{\rm XY} \Big( \ket i \bra i_{\rm X} \otimes I_{\rm Y} \Big) \ket{k}_{\rm  X} \nonumber \\ 
&& = \sum_{k} \Big( \bra k_{\rm X} \otimes I_{\rm Y} \Big) \rho_{\rm XY} \Big( \ket k_{\rm X} \otimes I_{\rm Y} \Big) 
= \rho_{\rm Y}, 
\label{rhopp}
\ee 
where the second-to-last equation is the known definition of the partial X-trace of $\rho_{\rm XY}$.

If Victor makes the projection (\ref{projV}) to get his pairs entangled, Alice and Bob need a classical signal from Victor to realize that their particles are entangled. 
Without this classical signal, they must sum over all possible outcomes to describe the system with 
$\rho''_{\rm XY}$. As we have shown, this has no observable effect because the reduced density matrix of Alice's and Bob's part will not be changed (see (\ref{rhopp})).

\subsection*{Annihilation with pure state particles}
For completeness, we also study the case in which the particle inside the black hole is not entangled with any other subsystem (we call this particle ``A''). Therefore, the state (\ref{BHp}) is replaced with
 \be 
\ket{\phi_2} = \sum_{\omega'} f(\omega') \ket{\omega'}_{\rm A } \, .
\label{BHp2B}
\ee
An analysis similar to the one in (\ref{FIN}) gives the following final state $| f^{\prime \prime} \rangle$, 
\be
| f^{\prime \prime} \rangle & = & \mathcal N  \sum_{ \omega, \omega'}  f( \omega' ) e^{-\frac{  \pi \omega} { \kappa}} \ket{ \omega} ^{\rm out}  \ket{\omega'  -\omega}^{\rm int}  \ket{0}_{\rm A}   \nonumber \\
&=& \mathcal N  \sum_{\omega}  f( \omega ) e^{-\frac{  \pi \omega} { \kappa}} \ket{ \omega} ^{\rm out}  \ket{0}^{\rm int}  \ket{0}_{\rm A} \nonumber \\
&&   + \mathcal N  \sum_{\omega' \neq \omega}  f( \omega' ) e^{-\frac{  \pi \omega} { \kappa}} \ket{ \omega} ^{\rm out}  \ket{\omega'  -\omega}^{\rm int}  \ket{0}_{\rm A} \nonumber \\
&\equiv &| f^{\prime \prime} \rangle_{\rm (i) } + | f^{\prime\prime} \rangle_{\rm (ii) } \, .
\label{FIN3}
\ee
Assuming full annihilation, we end up with the pure state $| f^{\prime \prime} \rangle_{\rm (i) }$. Indeed, the initial non-entangled pure state has evolved into a non-entangled pure state as well. Only in the intermediate stage is the created Hawking pair entangled.

 \subsection*{Annihilation with general 2-particle state}
For the analysis developed in the preceding section, we assumed the 
particles inside the black hole to have the same energy, but it is straightforward to generalize to an arbitrary entangled state. 
We again consider a Hawking pair in the state (\ref{Main2}), and a particle pair inside the black hole in the state $\ket{\chi}$ defined as: 
\be 
\ket{\chi} = \sum_{\omega'} f(\omega') \ket{g(\omega')}_{\rm A} \ket{\omega'}_{\rm B} \, .
\label{bipartite}
\ee
A pure bipartite entangled state can always be written in this form, where $g(\omega')$ is a general function of its argument. The initial state (\ref{IN}) is replaced with 
$| i_g \rangle = | \psi \rangle \otimes | \chi \rangle$ and, if we assume the negative energy particle interacts with particle B, the resulting state is:

\be 
 \!\!\!\! && \ket{f_g} =\mathcal N \!  \sum_{\omega', \omega}  \!f(\omega') e^{-\frac{  \pi \omega} { \kappa} } 
\ket{\omega}^{\rm out} \ket{g(\omega')}_{\rm A} \ket{\omega' - \omega}^{\rm int}_{\rm B} \, . 
  \label{General-1}
\ee
Then after another pair is created, and the resulting state will be:
\be \ket{f_g} && \to  \mathcal N~^2 \sum_{\omega', \omega, \omega''}  f(\omega') e^{-\frac{  \pi \omega} { \kappa} } e^{-\frac{  \pi \omega''} { \kappa''} } 
\ket{\omega}^{\rm out} \ket{\omega''}^{\rm out} \nonumber \\
&& \otimes \ket{g(\omega') - \omega''}_{\rm A} \ket{\omega' - \omega}^{\rm int}_{\rm B} .
\ee

Now as usual, upon evaporation, the state becomes 
\be \!\!\!\! \!\! \! \!  \to  \mathcal N~^{\!\! 2} \!   \sum_{\omega}  f(\omega) e^{-\frac{  \pi \omega} { \kappa''(\omega)} \!  -\frac{  \pi \omega} { \kappa(\omega) } } 
\! \ket{ g(\omega)}^{\rm out} \!   \ket{\omega}^{\rm out} \!  \ket{0 }^{\rm int}_{\rm A}\!  \ket{0}^{\rm int}_{\rm B}\!  \! .
\label{General0} 
\ee
So regardless in what state the inside particles are, it will be swapped outside upon evaporation.

 \subsection*{Annihilation with general n-particle state}  \label{multipartite}
We now consider a general multipartite entangled pure state describing a black hole resulting from a gravitational collapse. For the sake of simplicity, here we do not consider initial mixed states. 
However, our analysis applies in that case too. 
This will also help us to understand the previously mentioned case in which the incident Hawking particle scatters inside the black hole to produce more than one particle.

The multipartite matter state is a generalization of the simple bipartite state given in (\ref{bipartite}), namely 
\be
\ket{\Psi} &=& \sum_{\omega_1, ...\omega_k} \! f(\omega_1, \dots ,\omega_k) | {\omega_1}\rangle_{\rm A_0} 
| g_1(\omega_1,...,\omega_k)\rangle_{\rm A_1}\nonumber \\
&&| g_2(\omega_1,...,\omega_k)\rangle_{\rm A_2}  \dots 
|g_k(\omega_1,...,\omega_k)\rangle_{\rm A_k} \, , 
\label{MultiMatter}
\ee
where $f(\omega_1,...,\omega_k)$ is a general phase factor and ${\rm A_0}, \dots, {\rm A_k}$ are $k+1$ particles. 
Now, consider an incident Hawking particle of energy $\omega$ that scatters with the particle $\rm A_0$ to produce a particle of energy $\omega_1 - \omega$.
The state of the entire system, before the interaction takes place, is the tensor product of (\ref{MultiMatter}) and (\ref{Main2}), namely, $\ket{\Psi'} \equiv \ket{\Psi} \otimes \ket{\psi}$, 
\be 
\hspace{0cm}
 \ket{\Psi'}  &&= \mathcal N \hspace{-0.3cm} \! \sum_{\omega_1, ...,  \omega_k,\omega}  \hspace{-0.3cm} f(\omega_1,\!...,\omega_k) 
\, {\rm e}^{-\frac{  \pi \omega} { \kappa}} | {\omega_1}\rangle_{\rm A_0}
|g_1(\omega_1,\!...,\omega_k)\rangle_{\rm A_1} \nonumber \\
&& |g_2(\omega_1,\!...,\omega_k)\rangle_{\rm A_2}  \dots 
|g_k(\omega_1,\! ...,\omega_k)\rangle_{\rm A_k} \! \otimes \ket{\omega}^{\rm int} \ket{\omega}^{\rm out}\!\! \nonumber .
\ee
When the ``int'' particle interacts with the particle $\rm A_0$,
the state becomes
\be 
 \hspace{-0.4cm}
\!\! \ket{\Psi''}  = \, && \mathcal{N}  \! \!\! \! \! \! \sum_{\omega_1, ...,\omega_k, \omega} \! \! \!\! \! f(\omega_1,\! ...,\omega_k) {\rm e}^{-\frac{  \pi \omega} { \kappa}} | {\omega_1}_{\rm A_0}  - \omega \rangle
| g_1(\omega_1,\! ...,\omega_k)\rangle_{\rm A_1} \nonumber \\
&&  | g_2(\omega_1,\! ...,\omega_k)\rangle_{\rm A_2} \dots 
\ket{g_k (\omega_1,\! ..., \omega_k)}_{\rm A_k}\! \! \otimes \ket{\omega}^{\rm out} .
\ee 
Therefore, the resulting particle of energy ${\omega_1}_{\rm A_0}  - \omega$ is entangled with the black hole matter and with the Hawking ``out'' particle as well.
If more Hawking pairs are created, we have more ``out'' particles entangled with the black hole matter and the state is:
\be 
&&\! \! \! \!\!\!\!\! \! \! \! \ket{\Psi^{(k)}}  =  \mathcal N^{k+1}  \!  \! \!  \!  \sum_{\omega_1,\dots , \omega^{(k)}   } \! \! \! \!   \! \!  f(\omega_1,...,\omega_k) {\rm e}^{- \pi (\frac{ \omega}{ \kappa}+ \frac{ \omega'}{ \kappa'}+ \frac{ \omega''}{ \kappa''} +...) }   
\ket{ {\omega_1}_{\rm A_0} \! \!   -  \! \omega }\nonumber \\
&&  \ket{  g_1(\omega_1,...,\omega_k)_{\rm A_1} - \!   \omega'} 
\dots | g_k(\omega_1,...,\omega_k)_{\rm A_k} - \! \omega^{(k)}\rangle_{\rm BH} \nonumber \\
&& \hspace{2 cm}  \otimes \ket{\omega, \omega' ,...,\omega^{(k)} }^{\rm out} 
\label{General},
\ee 
where the sum above is on all of the frequencies $\omega_1, \dots , \omega_k, \omega, \omega' , \omega'', \dots , \omega^{(k)}$.
We now have an entangled state involving all of the particles inside and outside.
If we assume full evaporation\footnote{This is equivalent to saying that an observer at infinity makes a measurement of the black hole mass.} of the black hole, the entanglement is swapped to the outside radiation and the state reads:
\be 
 \mathcal N^{k+1}\! \! \! \! \! \sum_{\omega_1,...\omega_k} \!\!\!\! f(\omega_1,...,\omega_k) && {\rm e}^{- \pi (\frac{ \omega_1}{ \kappa_1}+ \frac{ g_1}{ \kappa_{\rm g_1}}+ \frac{ g_2}{ \kappa_{\rm g_2} }+...)}  \!\ket{0}_{\rm BH}  \nonumber \\
&&  \! \!  \! \! \! \!  \! \! \! \!  \! \! \! \!  \! \! \! \!  \! \! \! \!  \! \! \! \!  \! \!  \! \!   \! \!   \! \!   \! \! \otimes 
\ket{ {\omega_1}}_{\rm A_0}^{\rm out}
\ket{ g_1(\omega_1,...,\omega_k)}_{\rm A_1}^{\rm out}
\ket{ g_2(\omega_1,...,\omega_k)}_{\rm A_2}^{\rm out} \dots \nonumber \\
&& \dots \ket{g_k(\omega_1, \dots,\omega_k)_{\rm A_k} }^{\rm out} \!\! ,
\label{finalState}
\ee
where we labeled the states with the index ${\rm A}_i$ as well to keep track of the ``int'' particles that have been annihilated with the particles ${\rm A}_1, \dots {\rm A}_k$.

The state (\ref{finalState}) is clearly an entangled pure state of Hawking's  ``out'' particles after the black hole has fully evaporated. 
Note that the state (\ref{General}) is a superposition of all the energy's eigenstates. Therefore, the projection to the particular final state (\ref{finalState}) is only due to black hole full evaporation and not to an intrinsic unitarity violation. 

The outcome of this section can be summarized as follows. The pure entangled state describing matter inside the the black hole (\ref{MultiMatter}) evolves into the pure entangled state at $I^+$ (\ref{finalState}). Here, we only assumed annihilation inside the black hole between negative and positive energy particles.

\end{document}